 \documentclass[final,3p,times,numbers,sort&compress]{elsarticle}

\usepackage{amssymb}
\usepackage{subfigure}

\usepackage[titletoc]{appendix}

\usepackage{multirow}
\usepackage{amsmath}
\usepackage{booktabs}
\usepackage{color}
\usepackage{subeqnarray}
\usepackage{bm}
\usepackage{color}
\usepackage[T1]{fontenc}

\date{}

\begin{document}

\begin{frontmatter}



\title{On the conjugate interface conditions and Galilean invariance}


\author[bjtu]{Yang Hu\corref{cor1}}
\ead{yanghu@bjtu.edu.cn}

\cortext[cor1]{Corresponding author}

\address[bjtu]{School of Mechanical, Electronic and Control engineering, Beijing Jiaotong University, Beijing, P.R.China 100044}

\begin{abstract}
In the referred paper("H. Karani, C. Huber, Physical Review E, 91(2)(2015) 023304"), a total heat flux continuity condition for conjugate heat transfer problems with moving interfaces was proposed. 
The authors asserted both conductive and advective heat fluxes are conserved simultaneously in their formulation. This condition
had been cited by many subsequent studies. However, it is found that the total heat flux continuity condition violates Galilean invariance. 
The original diffusion heat flux continuity condition is reasonable for both stationary and moving interfaces.   

\end{abstract}


\end{frontmatter}



\section{\label{sec:level1}Introduction}

\begin{figure}[!ht]
  \begin{center}
      \includegraphics[scale=0.4]{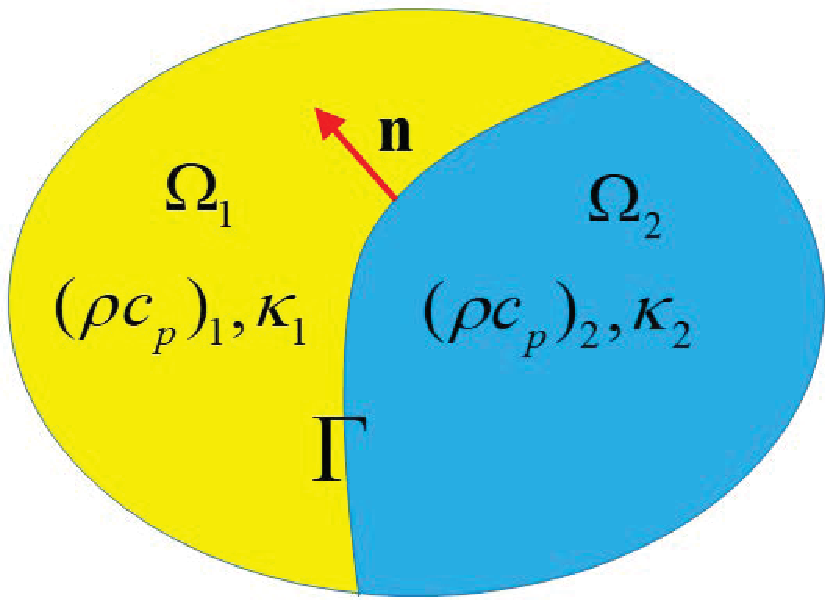} \hspace{1cm}
  \end{center}
  \caption{A schematic diagram of conjugate heat transfer problems.}
\label{multi-Domain}
\end{figure}

As shown in Fig. \ref{multi-Domain}, heat transfer in real world usually occurs in multi-domain or multiphase systems, known as 
the conjugate heat transfer problems \cite{John:2019,Zhang:2023}. The governing equation of the temperature field
in entire domain $\Omega_1 \cup \Omega_2$ is written as 
\begin{eqnarray}\label{CDE}
  \frac{\partial (\rho c_p T)}{\partial t}+\nabla \cdot (\rho c_p \textbf{u}T)=\nabla \cdot (\kappa \nabla T),
\end{eqnarray}
where $\rho$, $c_p$ and $\kappa$ are the density, specific heat capacity and thermal conductivity, respectively. These physical parameters
can be different in subdomains $\Omega_1$ and $\Omega_2$. For the rigid or incompressible media, the fluid velocity vector $\bm{u}$ 
satisfies the divergence-free condition. To obtain a well posed model for temperature field, 
the so called conjugate interface conditions should be given \cite{Perelman:1961,Hu:2015camwa,Hu:2015pre}. For the stationary interface, based on the conservation law, 
the temperature and normal heat flux on the interface $\Gamma$ satisfy the following continuity conditions 
\begin{eqnarray}
  &&T_1=T_2, \label{Temperature}\\
  &&\kappa_1 \nabla T_1 \cdot \textbf{n}=\kappa_2 \nabla T_2 \cdot \textbf{n},  \label{HeatFlux}
\end{eqnarray}
where $\textbf{n}$ is the unit normal vector on $\Gamma$. 

When we consider the conjugate heat transfer problems with moving interface, such as the particulate flows with heat transfer \cite{Yu:2006,Hu:2018}  or thermocapillary flow \cite{Young:1959,Hu:2019,Kim:1989}, 
the interface conditions on the evolving interface should be determined. However, there is a controversial about this issue. In many previous
papers \cite{Yu:2006,Hu:2018,Young:1959,Hu:2019,Kim:1989}, the interface conditions for moving interface share the same formulations with those for stationary interface, i.e. Eqs. (\ref{Temperature}) and (\ref{HeatFlux}).
Alternatively, Karani and Huber \cite{Karani:2015} argued that only diffusion heat flux continuity is applied in Eq. (\ref{HeatFlux}). For the moving interface, 
both diffusion heat flux $\kappa \nabla T$ and advective heat flux $\rho c_p \textbf{u} T$ should be considered. As a result, the following total heat flux continuity condition hold 
\begin{eqnarray}\label{TotalHeatFlux}
  (-\kappa \nabla T+\rho c_p \textbf{u} T)_1 \cdot \textbf{n}=(-\kappa \nabla T+\rho c_p \textbf{u} T)_2 \cdot \textbf{n}.
\end{eqnarray}
The above condition (\ref{TotalHeatFlux}) has been used and mentioned in many literatures \cite{Yue:2021,Wang:2019,Gao:2017,Ji:2020,Korba:2020,Karani:2017,Xu:2017,Wu:2017,Reynolds:2018,Wang:2022,Majumder:2024}. 

It is necessary to clarify the rationality of two kinds of interface conditions (Eqs. (\ref{HeatFlux}) and (\ref{TotalHeatFlux})). 
In this comment, the Galilean invariance in conjugate heat transfer problem is studied.
The forms of the temperature field equation and two types of interface conditions in two inertial frames are checked. 
We prove the total heat flux continuity condition does not obey the Galilean invariance. Moreover, through analysis of thermal equilibrium 
in a one dimensional infinite moving rod, and conjugate heat transfer in a channel with a vertical interface between both components, 
the irrationality of the total heat flux continuity condition is further confirmed. 

\section{Galilean invariance test}

We consider two inertial frames $S:(\textbf{x},t)$ and $\bar{S}:(\bar{\textbf{x}},\bar{t})$, and the following variables transformation is used 
\begin{eqnarray}
  &&\bar{\textbf{x}}=\textbf{x}+\textbf{U}_0 t, \\
  &&\bar{t}=t,\\
  &&\bar{\textbf{u}}=\textbf{u}+\textbf{U}_0,
\end{eqnarray}
where $\textbf{U}_0$ is a constant velocity vector. The derivatives of time and space have the following relations
\begin{eqnarray}
  &&\nabla=\bar{\nabla}, \\
  &&\frac{\partial }{\partial t}=\frac{\partial }{\partial \bar{t}}+\textbf{U}_0 \cdot \bar{\nabla}.
\end{eqnarray}
For the scalar quantity $T$ and the geometric quantity $\textbf{n}$, we have 
\begin{eqnarray}
  &&\bar{T}(\bar{\textbf{x}},\bar{t})=T(\textbf{x},t), \\
  &&\bar{\textbf{n}}(\bar{\textbf{x}},\bar{t})=\textbf{n}(\textbf{x},t).
\end{eqnarray}
Then the time derivative term, convective term and diffusion term in Eq. (\ref{CDE}) can be transformed as  
\begin{eqnarray}
  &&\frac{\partial (\rho c_p T )}{\partial t}=\frac{\partial (\rho c_p \bar{T})}{\partial \bar{t}}+\textbf{U}_0 \cdot \bar{\nabla}(\rho c_p \bar{T})\\ \nonumber
  &&=\frac{\partial (\rho c_p \bar{T})}{\partial \bar{t}}+ \bar{\nabla}\cdot(\rho c_p \textbf{U}_0\bar{T}),\\
  &&\nabla \cdot (\rho c_p \textbf{u}T )=\bar{\nabla} \cdot [\rho c_p (\bar{\textbf{u}}-\textbf{U}_0)\bar{T}], \\ 
  &&\nabla \cdot (\kappa \nabla T)=\bar{\nabla} \cdot (\kappa \bar{\nabla} \bar{T}). 
\end{eqnarray}
One can obtain
\begin{eqnarray}
  \frac{\partial (\rho c_p \bar{T})}{\partial \bar{t}}+ \bar{\nabla}\cdot(\rho c_p \bar{\textbf{u}}\bar{T})=\bar{\nabla} \cdot (\kappa \bar{\nabla} \bar{T}). 
\end{eqnarray}
It denotes that the temperature field equations in inertial frames $S$ and $\bar{S}$ have the same form. This is a trivial conclusion.
It is easy to know the time derivative term and convective term should be tied-up to keep Galilean invariance. In fact, 
the material derivative or total derivative should be treated as a whole according to the Lagrangian description and Eulerian description. 
The focus of this letter is to check the heat flux continuity condition using Galilean invariance test. we can see that the total heat flux 
continuity condition in $\bar{S}$ is  
\begin{eqnarray}\label{TotalHeatFlux-g}
  &&[-\kappa \bar{\nabla} \bar{T}+\rho c_p (\bar{\textbf{u}}-\textbf{U}_0) \bar{T}]_1 \cdot \bar{\textbf{n}} \\ \nonumber
  &&=[-\kappa \bar{\nabla} \bar{T}+\rho c_p (\bar{\textbf{u}}-\textbf{U}_0) \bar{T}]_2 \cdot \bar{\textbf{n}}.
\end{eqnarray}
It means
\begin{eqnarray}
  &&(-\kappa \bar{\nabla} \bar{T}+\rho c_p \bar{\textbf{u}} T)_1 \cdot \bar{\textbf{n}}\\ \nonumber
  &&=(-\kappa \bar{\nabla} \bar{T}+\rho c_p \bar{\textbf{u}} T)_2 \cdot \bar{\textbf{n}}+[(\rho c_p)_1-(\rho c_p)_2] \bar{T} \textbf{U}_0 \cdot \bar{\textbf{n}}.
\end{eqnarray}
Obviously, an addition term $[(\rho c_p)_1-(\rho c_p)_2]\bar{T}\textbf{U}_0 \cdot \bar{\textbf{n}}$ appears. The total heat flux continuity condition 
could not satisfy Galilean invariant. On the other hand, for the diffusion flux continuity condition, it is easy to know
\begin{eqnarray}
  \kappa_1 \bar{\nabla} \bar{T}_1 \cdot \bar{\textbf{n}}
  =\kappa_2 \bar{\nabla} \bar{T}_2 \cdot \bar{\textbf{n}}.
\end{eqnarray}
The Galilean invariance is kept. From a physical standpoint, the total heat flux continuity condition is unreasonable.

\begin{figure}[!ht]
  \begin{center}
      \includegraphics[scale=0.4]{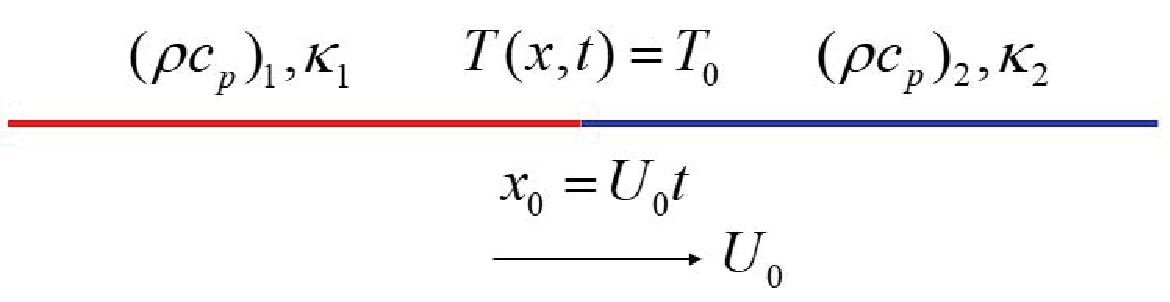} \hspace{1cm}
  \end{center}
  \caption{Thermal equilibrium in a one dimensional infinite moving rod.}
\label{one-dimensionalcase}
\end{figure}

\begin{figure}[!ht]
  \begin{center}
      \includegraphics[scale=0.4]{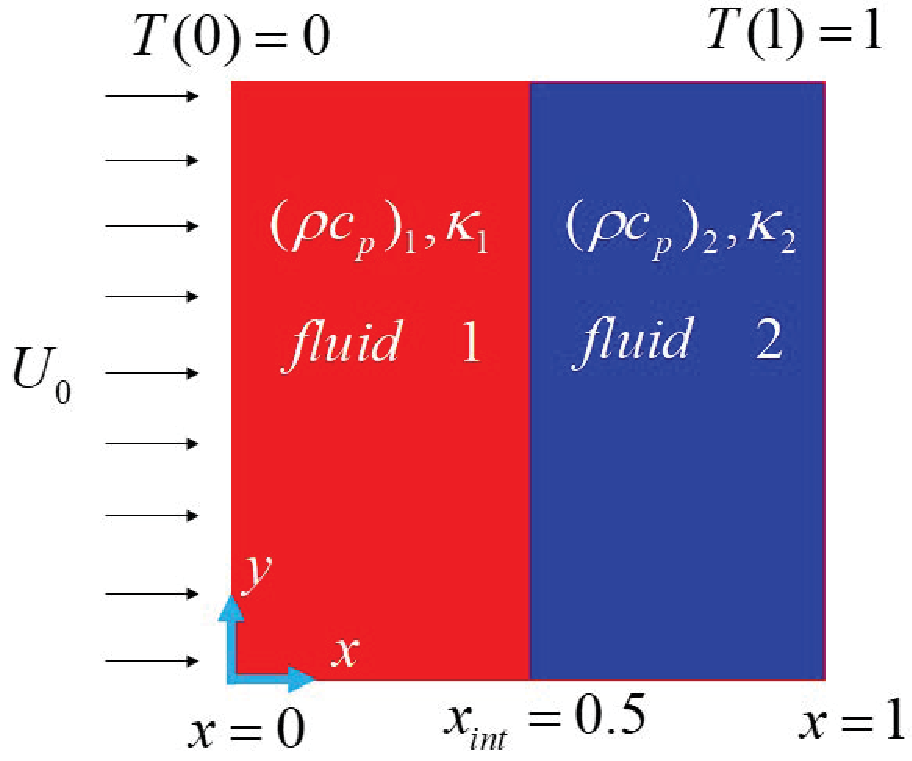} \hspace{1cm}
  \end{center}
  \caption{A schematic of conjugate heat transfer in a channel with a vertical interface between both components.}
\label{Steadyvertical}
\end{figure}

For better explanation, a simple example is discussed. As shown in Fig. \ref{one-dimensionalcase}, we consider the thermal equilibrium in a one dimensional infinite rod which is composed of 
two solids with different thermophysical properties. The rod moves uniformly with a velocity $U_0 \neq 0$.  Because there is no heat source,
the rod will be at the same temperature $T_0$ when the thermal equilibrium state is reached. At the interface point $x_0=U_0 t$, we notice that
the total heat flux continuity condition gives
\begin{eqnarray}
    (\rho c_p)_1 U_0 T_1-\kappa_1 \frac{dT_1}{dx}=(\rho c_p)_2 U_0 T_2-\kappa_2 \frac{dT_2}{dx}.
\end{eqnarray}
It leads to $(\rho c_p)_1=(\rho c_p)_2$, which is conflicted with our assumption. However, it is obvious that the diffusion flux continuity condition is valid.

Moreover, in the work of Karani and Huber\cite{Karani:2015}, we notice that an analytical solution for conjugate heat transfer in a channel with the vertical interface between both
components is derived to support the total heat flux continuity condition. The similar approach can be seen also in \cite{Yue:2021}. 
The geometric and physical configurations of this example is shown in Fig. \ref{Steadyvertical}.
This analytical solution has been used to check the numerical accuracy of some numerical schemes and 
the correctness of total heat flux continuity condition. It should be pointed out that there is an unreasonable setting in this case.
As we know, the phase interface should be driven by the velocity field and its velocity should be equal to value of velocity field on the interface.
 However, we can find that the phase interface $x_{int}$ between two fluids remains motionless under the velocity field. In other word, 
 $x_{int}$ is a constant. It means the term $\nabla \cdot (\rho c_p \textbf{u} T)$ partially loses the meaning of convection. 
 This treatment breaks the objectivity of material derivative and Galilean invariance. 
 In conclusion, this case is a pure mathematical problem, but not a real world problem.

\end{document}